\documentclass[letterpaper,useAMS,usenatbib]{mn2e}
\usepackage{comment}
\usepackage{amssymb}   
\usepackage{amsmath}   
\usepackage{amssymb}   
\usepackage{bm}

\usepackage{xspace}
\usepackage{graphicx}
\usepackage{verbatim}

\usepackage[T1]{fontenc}
\usepackage{aecompl}

\newcommand{\be}{\begin{equation}}
\newcommand{\ee}{\end{equation}}

\title[Missing Satellites in 3D]{The Missing Satellite Problem in 3D}
\author[Nierenberg et al.]{
A.~M.~Nierenberg$^{1, 2, 3}$, 
T.~Treu$^{4,5}$, 
N. ~Menci$^6$,
Y.~ Lu$^7$,
Paul Torrey$^{8, 9}$, 
M. ~Vogelsberger$^8$
\medskip\\
$^1$ Center for Cosmology and AstroParticle Physics, The Ohio State University, Columbus OH 43204, USA \\
$^2$ {nierenberg.1@osu.edu}\\
$^3$ CCAPP Fellow \\ 
$^4$ UCLA Physics \& Astronomy, 475 Portola Plaza, Los Angeles, CA 90095-1547, USA\\
$^5$ Packard Fellow\\
$^6$ NAF - Osservatorio Astronomico di Roma, via di Frascati 33, I-00040 Monteporzio, Italy\\ 
$^7$ Carnegie Observatories, 813 Santa Barbara Street, Pasadena, California, 91101, USA \\
$^8$ MIT Kavli Institute for Astrophysics and Space Research, 77 Massachusetts Ave. 37-241, Cambridge MA 02139,USA\\ 
$^9$ California Institute of Technology, Pasadena, CA 91125, USA\\
}

\def\be{\begin{equation}}
\def\ee{\end{equation}}

\def\rpower{\gamma_{\rm p}}

\def\alphas{\alpha_{\rm s}}

\def\dm{\delta_{\rm{m}}}
\def\dmo{\delta_{\rm{m,o}}}
\def\dmsat{\delta_{\rm{m,sat}}}

\def\dmo{\delta_{\rm{m,o}}}

\begin{document}
\date{Submitted to MNRAS}
\pagerange{\pageref{firstpage}--\pageref{lastpage}}\pubyear{2016}

\maketitle           

\label{firstpage}

                      
\begin{abstract}
It is widely believed that the large discrepancy between the observed number of satellite galaxies and the predicted number of dark subhalos can be resolved via a variety of baryonic effects which suppress star formation in low mass halos.
Supporting this hypothesis, numerous high resolution simulations with star formation, and associated feedback have been shown to reproduce the satellite luminosity function around Milky Way-mass simulated galaxies at redshift zero. However, a more stringent test of these models is their ability to simultaneously match the satellite luminosity functions of a range of host halo masses and redshifts. In this work we measure the luminosity function of faint (sub-Small Magellanic Cloud luminosity) satellites around hosts with stellar masses 10.5$<$Log$_{10}$M$_*$/M$_\odot<11.5$ to an unprecedented redshift of 1.5. This new measurement of the satellite luminosity function provides powerful new constraining power; we compare these results with predictions from four different simulations and show that although the models perform similarly over-all, no one model reproduces the satellite luminosity function reliably at all redshifts and host stellar masses. This result highlights the continued need for improvement in understanding the fundamental physics that governs satellite galaxy evolution. 

\end{abstract}

\setcounter{footnote}{1}

\section{Introduction}
\label{sec:intro}

One of the significant discrepancies between $\Lambda$~cold dark matter (CDM) and observation is the apparent lack of structure on sub-galactic scales. This `Missing Satellite Problem' \citep{Klypin++99, Moore++1999}, was first observed in the Local Group and has subsequently been observed in the mass function of isolated field galaxies \citep{Papastergis++11}. 

One commonly adopted solution to the missing satellite problem in the context of CDM, is that there are a large number of low mass dark matter subhalos that do not have detectable stars or gas.  The key to this solution is developing a comprehensive understanding of star formation in low mass halos. Simulations are now able to invoke a variety of mechanisms including UV heating during reionization, super-nova feedback, stellar winds, and tidal and ram pressure stripping by the host halo in order to reproduce the observed luminosity function of Milky Way satellites down to low masses at redshift zero \citep[e.g][]{Thoul++1996,Gnedin++00,Kaufmann++08, Maccio++11, Springel++10,Guo++11, Zolotov++12, Brooks++13, Starkenburg++13, Wetzel++16}. Presently, the baryonic solution to the missing satellite problem around Milky Way-like galaxies is not unique, with a variety of models all matching the data reasonably well. Much more information can be gained by comparing with observations of satellites around host galaxies of varying stellar mass and at a range of redshifts. Such a comparison can help distinguish between the roles of environmental and internal processes in regulating star formation in satellite galaxies.

It is also possible that the missing satellite problem is due to incorrect assumptions about dark matter. For instance, if dark matter is warm or significantly self-interacting, small scale structure can be erased at early times either due to finite particle velocities in the former case \citep[e.g.][and references therein]{Colombi++96}, or via a combination of dark sector Silk-damping and acoustic oscillations in the latter \citep{Vogelsberger++15, Cyr-Racine++15}. It is interesting to explore how different dark matter scenarios affect the star formation physics required to reproduce the observed luminosity function relative to the cold dark matter case.

Recent observations have generated a wealth of new information about satellite galaxies at low redshift, finding a significant dependence between the satellite luminosity function and host stellar mass and color \citep[e.g.][]{Guo++11,Liu++11,Lares++11,Strigari++12, Wang++12, Nierenberg++12, Ruiz++15, Sales++15, Lan++15}. Simulations have had varying success at reproducing these trends. For instance, \citet{Wang++12} found that the \citet{GuoQi++11} semi-analytic model applied to Millenium II halos matched some of the trend of increasing satellite numbers with host stellar mass, but significantly over-predicted the number of satellites around $\sim$10$^{11.5}$M$_\odot$ hosts relative to observations, as well as the number of bright satellites around lower stellar mass hosts. 

In addition to host stellar mass and morphology, redshift provides another important dimension along which to constrain star formation models, as different star formation processes occur on a broad range of time scales. Recent studies have begun to measure the number of satellites around higher redshift hosts between z of $0.4-2$ \citep{Newman++11, Nierenberg++12, Tal++14}. \citet[][(hereafter N13)]{Nierenberg++13} compared the measurement of the satellite luminosity function in bins of both redshift from 0.1-0.8 and host stellar mass to theoretical predictions from \citet{GuoQi++11}, \citet{Lu++12} and \citet{Menci++12}, finding that although all the models matched the luminosity function of Milky Way satellites, they each predicted significantly different luminosity functions in the other host mass and redshift regimes, demonstrating the power of comparing to observations outside of the Local Group.

In this work we extend our observation of the satellite luminosity function to fainter satellites and higher redshifts using deep, near-IR imaging from Cosmic Assembly Near-infrared Deep Extragalactic Legacy Survey \citep[CANDELS][]{Grogin++11, Koekemoer++11}. With the power of near-IR images, we are able to double our redshift baseline (from redshift 0.8 to 1.5) and reach an order of magnitude fainter satellites at z~0.8, with respect to our previous work. We combine the new measurement with lower redshift results from \citet[][(hereafter N12)]{Nierenberg++12}, and compare these results with four state-of-the-art simulations. The models are: 1) a CDM merger tree simulation with semi-analytic star formation\citep{Menci++14}, 2) a WDM merger tree with the same semi-analytic star formation implementation as in the CDM case also by \citep{Menci++14}, 3) a CDM $N$-body simulation with Bayesian-tuned semi-analytic star formation from \citet{Lu2014b} and 4) Illustris which is a CDM $N$-body simulation with hydrodynamical star formation \citep{Vogelsberger2014a}.

For the observational results, we assume a flat $\Lambda$CDM cosmology with
$h=0.7$ and $\Omega_{\rm m}=0.3$.  We note that although these values are slightly different from those used in each of the simulations, the variations caused by adjusting these parameters are much smaller than the measurement uncertainties. All magnitudes are given in the AB
system \citep{Oke++1974}.

\section{Data}
\label{sec:data}

To study the properties of satellites at a wide range of redshifts, we combine data from the COSMOS field \citep{Scov++07}and the CANDELS fields. The relatively wide COSMOS survey has $\sim$1.7 square degrees of HST F814 imaging, and is useful for constraining the satellite luminosity function at lower redshift and for higher stellar mass hosts. All results from COSMOS I814 imaging in this work are taken from N12 which contains a detailed discussion of the data that was used.

CANDELS has $\sim$0.25 square degrees of deep, near-infrared F160W, F140W and F125W HST imaging which enables the detection of fainter satellite galaxies at higher redshifts than COSMOS. To detect satellites in CANDELS, we make extensive use of data products provided by the 3D-HST team \citep{Brammer++12, Skelton++14}, which include reduced HST imaging as well as photometric, stellar mass and redshift catalogs. 

\subsection{Host galaxy selection}

We study satellites around host galaxies with stellar masses 10.5$<\log_{10}[$M$^*/$M$_\odot] <11.5$ between redshifts 0.1$<$z$<$1.5, using the stellar mass and redshift catalogs from the 3D-HST survey for CANDELS, and from \citet{Ilbert++09, Lilly++07} for COSMOS. The host stellar mass limits were selected to ensure sufficient signal of at least $\sim$ one satellite per host on the low mass end, and a well constrained halo-mass to stellar mass relation on the higher mass end. The latter restriction is important because we select satellites based on R$_{200}$ of the hosts. Where possible, spectroscopic redshifts were used, otherwise photometric redshifts were used from \citet{Ilbert++09} in COSMOS, and from 3D HST photometric catalogs in CANDELS for objects with photometric redshift quality indicator $Q_z<3$ as recommended by \citet{Brammer++08}.

We require hosts to be isolated, defined by being at least R$_{200}$ away from all galaxies more than twice as massive and at the same redshift $(z-z_{\rm{host}})/z_{\rm{host}}<0.007$.  We estimate R$_{200}$ from the stellar mass to halo mass relation from \citet{Dutton++10}.

In total, COSMOS contains 3038 host galaxies between redshifts $0.1-0.8$ which matched our criteria. Figure 1 of N12 shows the stellar mass distribution in bins of redshift for COSMOS hosts. The deeper and narrower CANDELS has 1708 host galaxies between redshifts 0.4 to 1.5. Figure 1 shows the stellar mass distribution of CANDELS hosts in bins of redshift.

\begin{figure}
\centering
\includegraphics[scale=0.5, trim = 25 0 250 0 clip = true]{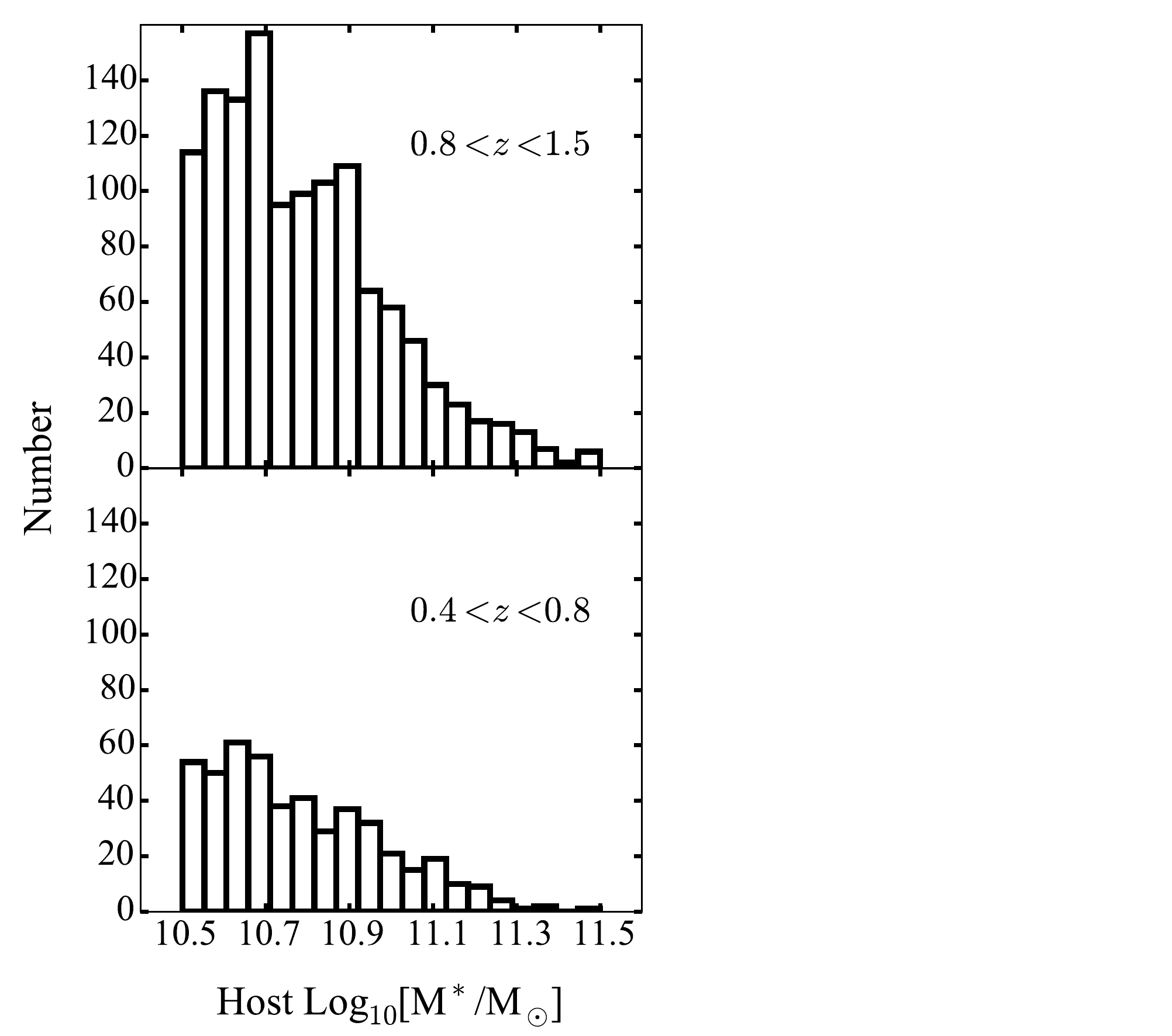}
\caption{The stellar mass distribution of host galaxies in CANDELS divided into bins of redshift. Figure 1 of N12 shows the corresponding distribution of host galaxies in COSMOS.}
\label{fig:prmass}
\end{figure}

\subsection{CANDELS Satellite candidate selection}

We use the 3D HST PSF-matched photometric catalogs to select the majority of satellite candidates. These catalogs were created with SourceExtractor \citep[][hereafter SE]{Bertin++1996} on PSF matched, variance weighted mean combined F160W, F140W and F125W images. 
Satellite candidates include all non-stellar \footnote{SE {\tt CLASS\_STAR}$<0.8$ and 3D HST {\tt star\_flag}~!=1} objects brighter than F160W $<$ 25 AB magnitudes between  $0.07<$R/R$_{200,\rm{host}}<0.5$ where R$_{200,\rm{host}}$ is estimated based on the host stellar mass and stellar mass to halo mass relation from Equation 3 of \citet{Dutton++10} for early-type galaxies. The limit of F160W$<$25 magnitude is chosen to ensure over 90\% detection completeness in both the CANDELS deep and wide imaging based on Figure 14 from \citep{Skelton++14}, while the radial limits ensure accurate photometry near the host galaxies, and a favorable ratio of satellites to background/foreground objects in the outer region.

As we demonstrated in \citet[][hereafter, N11]{Nierenberg++11} and N12, even with deep, high resolution HST imaging, photometric accuracy and completeness suffer within several effective radii of a bright central galaxy. To counter this, we empirically model and subtract the host light profile, and then perform our own object detection and photometry in small regions near the host galaxies, using SE parameters which match those used to create the 3D HST catalogs. We then add the new object detections to the full 3D HST catalogs, and update the photometry only for the region near the host galaxies. This method improves completeness and photometric accuracy very near the host galaxies (N11, N12).
We perform the empirical host subtraction separately in both the detection image, as well as in the F160W science image. We then apply the same empirical photometric correction to the F160W MAG\_AUTOs as applied to make the 3DHST photometric catalogs \citep[see][Figure 10]{Skelton++14}.

\section{Statistical Analysis}
We detect satellites statistically in monochromatic F160W data in CANDELS (F814 for COSMOS), as an increase in the number density of objects relative to the background/foreground number density measured outside of the host galaxy R$_{200}$. 
The statistical model framework is described in detail in N11 and N12. The measurement is made in two subdivided bins of host stellar mass 10.5$<\log_{10}[$M$_*/$M$_\odot] <11$, 11$<\log_{10}[$M$_*/$M$_\odot] <11.5$ and three bins of redshift $0.1<\rm{z}<0.4$, $0.4<\rm{z}< 0.8 $, $0.8<\rm{z}< 1.5$. Each redshift and host stellar mass bin, is treated as an independent data set. Results for CANDELS and COSMOS are inferred separately. Model parameter definitions and priors are listed in Table 1 for CANDELS, and Table 1 in N12 for the COSMOS.

In brief, the number density of objects around the host galaxies is parametrised as the sum of a uniformly distributed number of background/foreground objects, and a population of satellite galaxies with a power-law spatial distribution with projected slope $\rpower$, such that $P(r_{\rm{sat}})\propto r_{\rm{sat}}^{\rpower}$. 

In this work, we update the model relative to N11 and N12, by directly inferring the slope and bright end cutoff of the satellite luminosity function rather than iteratively inferring the cumulative number of satellites brighter than a fixed $\delta_{\rm{m}} = \rm{m}_{\rm{sat}}  - \rm{m}_{\rm{host}}$, for a series of $\delta_{\rm m}$ values. 
We model the luminosity function of satellites in units of $\dm$ as a Schechter function with faint end slope $\alphas$ and bright end cutoff of $\dmo$:
\begin{equation}
\begin{array}{ll}
P(\dmsat|\alphas,\dmo) \propto & 10^{\frac{\alphas+1}{2.5}(\dmo-\dmsat)} \times\\
&\exp{[-10^{(\dmo-\dmsat)}]}
\end{array}
\end{equation}

The total model number of satellites per host, N$_{\rm{s,o}}$ between $0.07<$r/R$_{200}<0.5$ is defined between $\delta_{\rm{m,min,o}}$ and $\delta_{\rm{m,max,o}}$, the minimum and maximum values respectively where the luminosity function can be measured reliably for the majority of the sample. These values vary depending on the average magnitude of the hosts in each redshift bin, as $\delta_{\rm{m,max,o}}$ for a given host is $25-\rm{m_{host}}$, the satellite luminosity function of brighter hosts can be measured to lower values of $\delta_{\rm m}$. Table 1 gives the values of the limits $\delta_{\rm{m,min,o}}$ and $\delta_{\rm{dm,max,o}}$ selected for the inference in each host redshift and stellar mass bin.

We adopt priors on the satellite spatial distribution, the background/foreground number density, and the slope of the background/foreground luminosity function. In N12 we found that $\rpower = -1.1\pm 0.3$ over a wide range of redshifts, satellite luminosities and host stellar masses and morphologies, so for this work we adopt a Gaussian prior on $\rpower$ with mean -1.1 and standard deviation 0.3. As we show in N12, adopting this prior does not affect the mean inferred number of satellites, but it does decrease the uncertainty, and is useful for constraining the properties of satellites around the much smaller sample of CANDELS host galaxies. 

We constrain the properties of the background/foreground galaxy in annuli outside of the region where we detect satellite galaxies. This method has been shown to accurately characterise the density of both random line of sight structure, as well as correlated structure which is not gravitationally bound to the host galaxies \citep{Chen++08, Liu++11}. We select annuli between 1$<$R/R$_{200, \rm{host}}<2$ to measure both the background/foreground number density, $\Sigma_{\rm b}$, and the slope of the number counts $\alpha_{\rm{b}}$, where $\rm{N}_{\rm{b}}(m) \propto \Sigma_{\rm b}10^{\alpha_{\rm{b}}m}$. These values do not depend strongly on the choice of the annuli radius, given the measurement uncertainty. We include $\Sigma_{\rm b}$ and $\alpha_{\rm b}$ as model parameters with Gaussian priors with mean and standard deviations given by the values measured in the outer annuli. This enables us to directly account for the effects of measurement uncertainties in $\Sigma_{\rm b}$ and $\alpha_{\rm b}$ in the inferred satellite properties. The prior values for these parameters are listed in Table 1. 

\begin{table*}
\begin{tabular}{lll}
\hline \hline
Parameter & Description & Prior  \\
\hline
N$_{\rm{s,o}}$ & Integrated number of satellites with $\delta_{\rm{m,min,o}}< \delta_{\rm{m}}< \delta_{\rm{m,max,o}}$ & U(0,100)$^{\rm{a}}$ \\
$\alphas$  & Faint end slope of the satellite luminosity function (Equation 1) & U(-2.9, 0) \\
$\dmo $    & Bright cutoff of the satellite luminosity function (Equation 1) & U(-8, 4) \\
$\rpower $& Projected satellite number density radial profile & G(-1.1, 3)$^{\rm{b}}$ \\
$\Sigma_{\rm{b}}$ & Number of background/foreground galaxies per arcmin$^2$ & Gaussian, varies with host mass, redshift and field. \\
$\alpha_{\rm{b}}$ & Slope of background/foreground magnitude number counts & G(0.28,0.01) \\
\hline\hline
\end{tabular}
\caption{Model parameter definitions and priors. (a) U(min,max) is a uniform prior between the min and max values. (b) G(mean, std) is a Gaussian prior defined by the mean and standard deviation.}
\end{table*}

\section{Results}

The inferred median values and 1$\sigma$ uncertainties for the satellite luminosity function parameters in CANDELS are given in Table 2 for the four host stellar mass and redshift bins. For all host stellar mass and redshift regimes, the prior values for the projected number density radial profile of satellites $\gamma_{\rm{p}}$, the number density of foreground/background objects and the slope of the background/foreground number counts were recovered, with no tightening of the constraints. This was expected given that the prior on $\rpower$ is derived from the much larger COSMOS sample, while the background/foreground number density is measured in a much larger region than the region in which satellite properties are inferred. Figure 2 shows the inferred cumulative satellite luminosity function for CANDELS (dark hatched region) and COSMOS (black points) with satellite numbers extrapolated to be between $0.07<$r/R$_{200,\rm{host}}<1$ based on the inferred value of the slope of the satellite number density radial profile $\gamma_{\rm{p}}$. 

For CANDELS we plot in black, regions which do not overlap with COSMOS, and which had a significant number of hosts ($>10$) with $\delta_{\rm{m, max, i}} = 25-m_{\rm{host}}$, to ensure that the luminosity function is well constrained by the data. In light grey we show the extrapolation of the luminosity function to fainter values of $\delta_{\rm{m}}$ where the data is less constraining, which is reflected in the increased uncertainty in the inferred luminosity function at these values. We also extrapolate the CANDELS results to brighter values of $\delta_{\rm{m}}$ to demonstrate the consistency with COSMOS results, and to show the overall shape of the luminosity function in the higher redshift bin. The number of satellites is well constrained and consistent with prior measurements of the luminosity function from COSMOS. N12 gives a detailed comparison between the lower redshift (0.1$<$z$<0.4$) COSMOS results and other results from the literature.

The inferred slope of the satellite luminosity function is $\alpha_{\rm{s}}\sim -1 \pm 0.5$, consistent within the measurement uncertainties for all the redshift and host stellar mass regimes. The bright cutoff, $\dmo$ is approximately $1\pm1$, again consistent between the four regimes.

This result is broadly consistent within the measurement uncertainties with the low redshift z$\sim 0.01-0.05$ luminosity function measurement of satellites brighter than M$_{r}\sim-14$ from \citet{Lan++15} which fit a double Schechter function to their satellites and found a slope of $\sim -1$ for satellites with absolute magnitudes brighter than -18 (corresponding to $\dm$ values of $\sim 2-3$ depending on host stellar mass), and slope of $\sim -1.6$ for satellites fainter than -18. 
Our result is also consistent with measurements of the slope of the stellar mass function of field galaxies and satellites between redshifts $\sim 0-4$ which typically find values of $\alpha$ between -1 to -2 \citep[e.g.]{Grazian++15, Duncan++14, Baldry++12, Santini++12, Graus++16, Lan++15}. Our data is not constraining enough to test whether satellite galaxies follow the field galaxy trend of decreasing $\alpha$ with increasing redshift.

It is interesting to compare these new high redshift satellite observations with local observations of Milky Way mass hosts.
Assuming the Milky Way has approximately doubled in stellar mass since z$\sim1$ \citep[e.g.][]{Behroozi++10, Lehnert++14}, Milky Way progenitors fall on the lower end of the lower stellar mass hosts in the redshift bin 0.8-1.5. Hosts in this stellar mass bin have $\sim 2\pm 0.7$ satellite galaxies with $\dm>4$, and $\sim 0.8\pm 0.3$ with $\dm>2$ corresponding to the present day magnitude offsets between the Small Magellanic Cloud (SMC) and the Milky Way, and the Large Magellanic Cloud (LMC) and the Milky Way respectively. This frequency is marginally higher, but still consistent within the measurement uncertainties, with low redshift observations where Milky Way stellar mass hosts have about 0.3$\pm$ 0.1 and 0.2 $\pm$ 0.1 SMCs and LMCs respectively \citep[see also][]{Liu++11, Guo++11, Lares++11}.

\begin{table}
\begin{tabular}{lllll}
\hline \hline
Log[M$_{\rm{host}}^*/$M$_\odot$] & z$_{\rm{host}}$ &$N_{\rm{s,o}}$ & $\alpha_s$ & $\dmo$ \\
\hline 
10.5-11.0  &  0.4-0.8 & 3$^{+1}_{-1}$ (3-6)  & -1.2$^{+0.4}_{-0.6}$          & $1^{+1}_{-1}  $\\
10.5-11.0 & 0.8-1.5 & 0.7$^{+0.6}_{-0.4}$ (2-4.5)   & -0.8$^{+0.5}_{-0.7}$          & 0.9$^{+1}_{-1}$  \\
11.0- 11.0 & 0.4-0.8 & 3$^{+3}_{-2}$ (2-5)  & -0.8$^{+0.5}_{-0.7}$          & 2$^{+1}_{-2}$  \\
11.0-11.5 & 0.8 -1.5 & 3$^{+1}_{-1}$ (2-5)  & -1.1$^{+0.6}_{-0.5}$          & 0.9$^{+1}_{-2}$  \\
\hline \hline
\end{tabular}
\caption{Inferred median and one sigma confidence intervals for the parameters defining the satellite luminosity function, in bins of host galaxy stellar mass and redshift. Parameters are defined in Table 1. Note that the normalisation of the luminosity function $N_{\rm{s,o}}$, is defined over different intervals of $\delta_{\rm{m}} = \rm{m_{sat}}-\rm{m_{host}}$ of the luminosity function for each of the bins, shown in parentheses to account for the fact that the completeness varies with redshift and host stellar mass. }\end{table}

\section{Comparison with theoretical models}

This new measurement of faint satellites provides a new constraint for models of galaxy formation. We compare our results with four theoretical predictions described below. Three of the predictions are from either semi-analytic or hydrodynamical models applied to cold dark matter cosmologies, and the fourth prediction is a semi-analytic model applied to a warm dark matter cosmology. 

\subsection{Menci}
The Menci model is a semi-analytic model applied to dark matter merging trees. We include a summary of the dark matter and semi-analytic models below, we refer the reader to \citet{Menci++14} for a detailed description. 

In this work we consider both a Cold and a Warm dark matter merging tree. Both models track subhalos after they have entered the virial radius of the main halo, enabling them to coalesce with the central galaxy after orbital energy loss due to dynamical friction, merge with another sub-halo, or survive as a satellite halo.
Cold Dark Matter halo merging trees are generated through a Monte Carlo procedure adopting the merging rates given by the Extended Press \& Schechter \citep[EPS, see][]{Lacey++93} formalism. 
The WDM power spectrum is generated using the method described by \citet{Menci++12}, with some modifications for the present work.  First, we have included a sharp-k filter to relate the power spectrum to the variance of density perturbations 
\citep[see][]{Benson++13,Schneider++13}. Secondly, we assume a thermal relic mass $m_X=1.5$ keV for the WDM candidate which yields a power spectrum corresponding to that produced by (non-thermal) sterile neutrinos with mass $m_{sterile}\approx 6-12$ keV, depending on the production mechanism \citep[see, e.g.][]{Kusenko++09, Destri++13}. Indeed, these constitute the simplest candidates \citep[see e.g.][]{Abazajian++14} for a dark matter interpretation of the origin of the recent unidentified X-ray line reported in stacked observations of X-ray clusters \citep{Bulbul++14, Boyarsky++14}. This particle mass is consistent with observational limits ($m_X\gtrsim 1.5$ keV) from high redshift galaxies \citet{Schultz++14} and ultra-faint dwarf galaxies \citep{Lovell++12, Lovell++14, Horiuchi++14}, although still in tension with Lyman-$\alpha$ forest constraints 
which yield $m_X\gtrsim 3.3$ keV \citep{Viel++13}. A discussion of the various uncertainties which may affect the constraints is given in \citet{Abazajian++11} and in \citet{Garzilli++15}.

The processes affecting baryons are connected in the same way to the evolution of both Cold and Warm dark matter halos. The baryonic model includes atomic cooling into rotationally supported discs following \citet{Mo++98}. Star formation occurs quiescently over long time scales ($\sim 1$ Gyr), and in bursts ($\sim 1$ Myr) triggered by galaxy interactions and disc instabilities. Star formation is suppressed via feedback from supernova and active galactic nuclei as described in \citet{Menci++08}.
The luminosity is computed by convolving the star formation histories of the galaxy progenitors with a synthetic spectral energy distribution, which we take from \citet{Bruzual++03} assuming a Salpeter IMF. 
The model includes tidal stripping of part of the stellar content of each satellite galaxy following \citet{Henriques++10}.

\subsection{Lu}

The Lu model is a semi-analytic model applied to a set of halo merger trees extracted from a large cosmological $N$-body simulation, the Bolshoi Planck simulation, which is same as the Bolshoi simulation \citep{Klypin2011a}, but with a cosmology favored by Planck data \citep{Planck2015} with parameters $\Omega_{\rm m,0}=0.30711$, $\Omega_{\rm \Lambda, 0}=0.69289$, $\Omega_{\rm b,0}=0.048$, $h=0.7$, $n=0.96$, and $\sigma_8=0.82$. The mass resolution of the simulation allows the model to track halos and subhalos with mass $\sim 7\times10^9 M_\odot h^{-1}$. The prescriptions for the baryonic processes implemented in the SAM are detailed in \citet{Lu2014b}. The model parameters governing star formation and feedback are tuned using an MCMC optimization to match the local galaxy stellar mass function \citep{Moustakas2013a}. Therefore, it is guaranteed to produce a global galaxy stellar mass function for the stellar mass range between $10^9$ and $10^{12}M_{\odot}$ at redshift zero within the observational uncertainty for the given parameterization of the model. 

\subsection{Illustris}

The Illustris simulation~\citep{Vogelsberger2014a} modeled a 106 Mpc$^3$ volume including both dark matter and baryons using the {\small AREPO} simulation code~\citep{Springel2010}.  
The Illustris model includes a range of physical processes (including radiative gas cooling, star formation, stellar feedback, and AGN feedback) that have been tuned to produce broad agreement with the cosmic star formation rate density and redshift $z=0$ stellar mass function~\citep{Vogelsberger2013, Torrey2014}.
For a full description of the Illustris simulation setup, see ~\citet{Vogelsberger2014b} and~\citet{Genel2014}.
The galaxy luminosities used here are assigned using the methods described in~\citet{Torrey2015a}.

\subsection{Comparison Results}

We compute the satellite luminosity function from the Lu and Illustris simulations by including all objects within R$_{200}$ of a host halo, where R$_{200}$ is estimated using the same stellar mass to halo mass relation as was applied to the observations \citep{Dutton++10}. We note that the number of objects within R$_{200}$ is significantly lower than the total number of subhalos identified by the FOF algorithm for both Illustris and Lu models. However, the number does not depend strongly on whether the \citet{Dutton++10} relation is used to estimate the host virial radius, or whether the simulation value of R$_{\rm{vir}}$ is directly used. For the Menci CDM and WDM merging trees, we directly use the simulation definition of satellite membership which is determined by the host virial radius.

We compute the $\chi^2$ between the model number of satellites per host and the measured numbers as a function of $\delta_{m}$, for discrete values of $\delta_{m}$ spaced in intervals of 0.5 for COSMOS data and 1 for CANDELS data. 
In regions where both CANDELS and COSMOS data overlapped, the COSMOS data is used in the comparison with theory owing to its higher precision. Although our data is composed of measurements in both F814W and F160W, the luminosity function of satellites as a function of  $\delta_{\rm{m}} = \rm{m_{sat}}-\rm{m_{host}}$ does not have a significant dependence on observed band in any of the theoretical models for redshifts $<0.8$.

We include the covariance between data points in our $\chi^2$ estimate, which is significant owing to the fact that we are considering the cumulative numbers of satellites, and thus the same host galaxy will contribute to the number of satellites in multiple bins.
The covariance matrix between data points was computed using 2500 draws from the Markov Chain Monte Carlo results for CANDELS while for COSMOS it was computed by bootstrap resampling the input data and re-inferring the satellite numbers. The black hatched region of the CANDELS measurement shows the range of CANDELS data which was used to compute the $\chi^2$.

\begin{table*}
\begin{tabular}{lllllll}
\hline \hline

z  &	Log$_{10}[$M$^{*}_{\rm {host}}/$M$_\odot$]& \# Data Points & Menci CDM&   Menci WDM & Lu & Illustris \\
 \hline
 $0.1-0.4$& 10.5- 11  & 13     & 17                &  18                   & 10          &  9 \\ 
& 11- 11.5                   &13      &14     		&  10		  &12           &  11 \\
 $0.4-0.8$ & 10.5- 11  & 9       &  6		& 11			  & 13          & 12 \\
&11- 11.5                    &11      &  31		& 9			  & 18	  & 20 \\
$0.8-1.5$ 	& 10.5- 11 &4        & 2  		&2			   &  5 	   &5 \\
& 11- 11.5                   &4        &   2	         &2			   & 3            & 2 \\
\hline
Total &                        &  54     &72	         &55	                     & 59           & 61 \\

\hline \hline
\end{tabular}
\caption{Model $\chi^2$ values, after accounting for covariance in the data, for each host stellar mass and redshift bin. We note that a continuous luminosity function is inferred CANDELS data, but to compare with predictions from simulations we sample the luminosity function values at discreet points spaced every 0.5 $\delta_{\rm{m}}$ within the dark hatched region of Figure 2. }
\end{table*}

Table 2 lists the $\chi^2$ values between the models and observed luminosity for each bin in host stellar mass and redshift, along with the corresponding number of data points. The $\chi^2$ values were 61, 72 and 59 for the Lu, Menci and Illustris CDM models, and 54 for the Menci WDM model for 53 degrees of freedom.  Accounting for the number of degrees of freedom, the $\chi^2$ goodness-of-fit test gives an approximately $\sim 20\%, 5\%, 15\%$ and $60\%$ chance that the data was drawn from each model respectively. These values are approximately equivalent, with only the Menci CDM model being somewhat disfavoured relative to the others, but at less than a 2$\sigma$ confidence, and the Menci WDM model performing the best.

\section{Discussion}

This data provides an important new constraint for models of satellite galaxy formation and evolution.  While overall there is not a strong preference for a particular model, it is interesting to consider how the models performed in the different regimes. In particular, the Menci WDM model provides the best overall match to the luminosity function around higher mass hosts while the Lu and Illustris models match the luminosity function of low-mass hosts at low redshifts. There is some tension between the numbers of faint satellites in the redshift range $0.4-0.8$ for lower stellar mass hosts, for all of the models, which somewhat underestimate the slope of the satellite luminosity function. Overall, models which produce accurate predictions for Milky Way mass hosts at low redshifts tend to underestimate the number satellite galaxies found around those hosts at higher redshifts and at higher stellar masses, and vice versa.  

This measurement provides complementary constraints to measurements of field galaxy properties. In particular, the Illustris and Menci CDM models both significantly over-predict the number of low mass $\log_{10}$M$_*$/M$_\odot<10$ field galaxies at redshift zero by a factor of $\sim 3-5$ \citep{Vogelsberger2014b, Genel2014, Menci++12}. While stronger feedback might resolve this discrepancy, in the case of Illustris, it would lead to a further suppression of the satellite luminosity function relative to observation. The Lu model matches the field luminosity function well at redshift zero but tends to under-predict the slope of the luminosity function at higher redshift \citep{Lu2014b}. As shown in \citet{Lu2014b} the model also underestimates the metallicities of low mass galaxies owing to the strong feedback implementation. 
One possible solution to these discrepancies in the context of a CDM cosmology may be in the form of a different feedback model such as the preventative feedback model of \citep{Lu++15}, in which star formation is suppressed due to the pre-heated intergalactic medium, rather than via ejective feedback processes such as outflows. This mode of suppression better reproduces the observed cold gas fractions, star formation histories and sizes of low mass galaxies for field galaxies, although it has yet to be tested for satellite galaxies. Preventative feedback may also impact the simulated mass-metallicity relation by reducing fraction of metals that are ejected from the galaxy~\citep{Zahid2014}. 

An interesting result of this comparison is that the variation between the predicted luminosity function in CDM models with different star formation models is as great as the variation between the CDM and WDM models with fixed star formation. The Menci WDM model provided the closest match to the data, but it under-predicted the slope of the satellite luminosity function at higher redshifts around the lower stellar mass host galaxies in a similar way to the CDM models. In a future work it would be interesting to compare observational results with a broader range of dark matter models such as those generated by ETHOS \citep{Vogelsberger++15, Cyr-Racine++15}. A direct measurement of the low mass halo mass function via gravitational lensing would provide a powerful constraint for these dark matter models \citep[][and references therein]{Treu++10} and therefore play an important role in constraining the star formation physics in low mass halos. In order to understand how luminous satellites populate the subhalos detected in gravitational lensing studies, it is important to study luminous satellites around gravitational-lens stellar mass hosts \citep[see also][]{Nierenberg++13b,  Fiacconi++16}.

\begin{figure*}
\centering
\includegraphics[scale=0.5, trim = 20 0 150 50, clip = true]{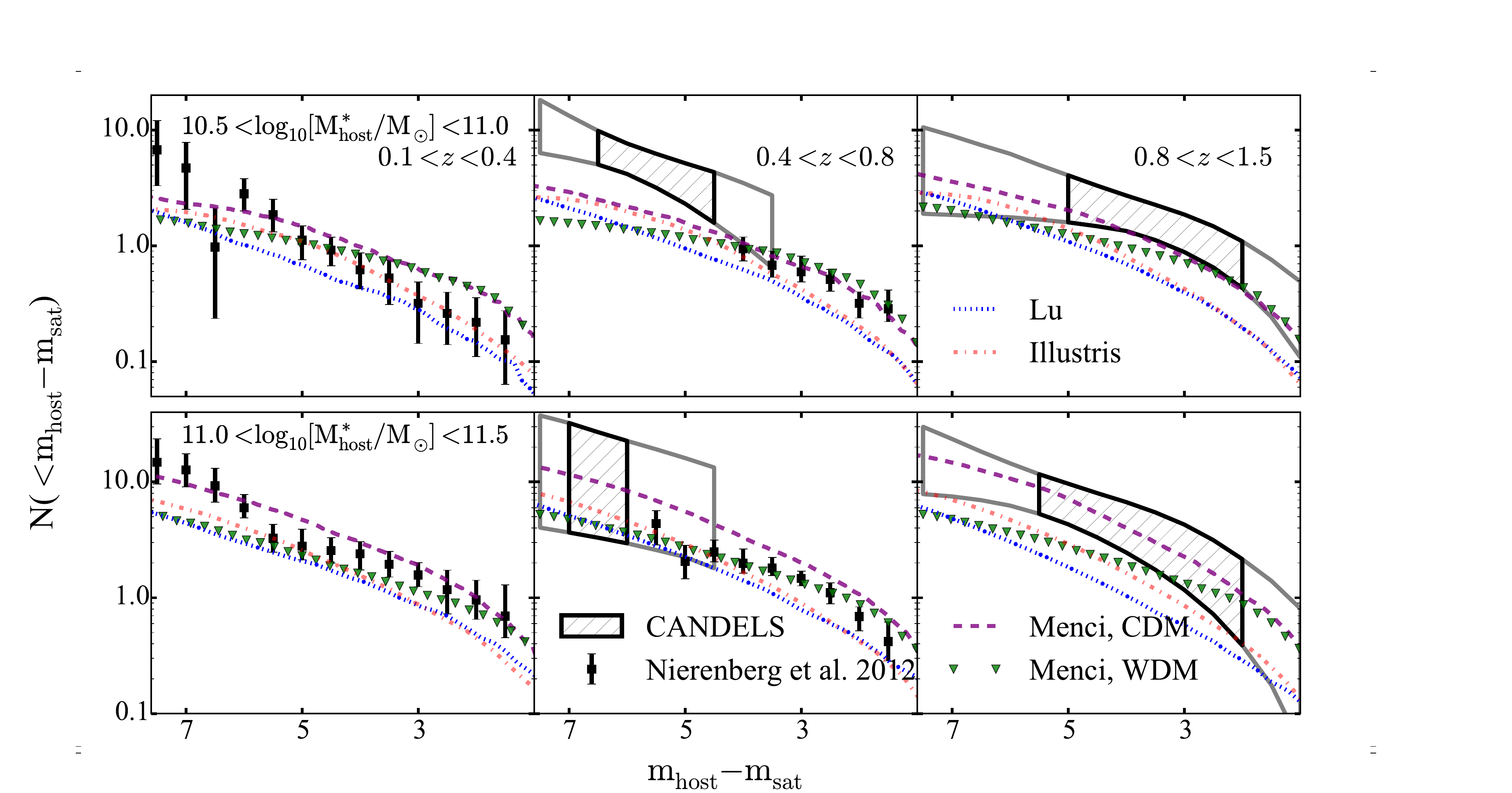}
\caption{The cumulative luminosity function of satellite galaxies of hosts divided by bins of redshift (from left to right) and stellar mass (from top to bottom) between 0.07$<$r/R$_{200}<1.0$. Black points with error bars and hatched region show the one sigma confidence interval from COSMOS F1814W and CANDELS F160W imaging respectively. The black points and hatched region were used to compute $\chi^2$ values for the comparison with theory, while the gray region represents extrapolations of the inferred luminosity function. Blue, purple and red lines lines are theoretical predictions with Cold Dark Matter from \citet{Lu2014b}, \citet{Menci++14}, and Illustris ~\citep{Vogelsberger2014a}. The green luminosity function is the same star formation model as for the Menci CDM model, but applied to a Warm Dark Matter halo mass function \citep{Menci++12}.}
\label{fig:prmass}
\end{figure*}

\section{Summary}

Using CANDELS F160W imaging, we measure the luminosity function of faint satellites around hosts with stellar masses between $10.5<$Log$_{10}$M$_*$/M$_\odot<11.5$ to a redshift of 1.5. The deep imaging enables us to detect satellites with $\dm=$m$_{\rm{sat}}$-m$_{\rm{host}}=4$ (fainter than the SMC) to an unprecedented redshift of 1.5, and to detect $\dm=7$ satellites between redshifts 0.4-0.8. We combine these new results with lower redshift (0.1-0.8) measurements from F814 COSMOS imaging from \citet{Nierenberg++12}. We compare these results to predictions from four theoretical models \citep{Menci++12, Menci++14, Lu2014b, Vogelsberger2014b}. While none of the models was significantly ruled out, different models matched the observations more or less well in different regimes of redshift and host stellar mass, which highlights the value of this data set in distinguishing between models which performed similarly at redshift zero around Milky-Way-mass host galaxies. 
This data provides important new constraining power for future implementations of feedback and dark matter physics in these models.

Our conclusions are summarised as follows:

\begin{enumerate}
 \item We infer the parameters of the satellite luminosity function faint end slope to be $\alpha_{\rm{s}}= -1\pm0.5$, and the bright cutoff relative to the host magnitude to be $\dmo = 1\pm1$, consistent over all redshift and stellar host stellar mass bins within the measurement uncertainties.
 
 \item We detect Small Magellanic Cloud-luminosity satellite galaxies to a redshift of 1.5, and find that Milky Way-like progenitors at redshift $\sim1$ have consistent numbers of LMC/SMC analogs with redshift zero galaxies. In general, the cumulative number of satellites per host as a function of the offset between host and satellite magnitude is constant as a function of redshift within the measurement uncertainties, but is higher for hosts with higher stellar masses.
 
 \item The theoretical models varied in their ability to predict the satellite luminosity function in different regimes of host stellar mass and redshift. Overall they performed similarly, however, models which fit the luminosity function of the satellites of low stellar mass hosts tended to under-predict the number of satellites around higher stellar mass hosts. Future predictions will need an adjusted implementation of stellar feedback as a function of host stellar mass and redshift in order to resolve this discrepancy. 
 
 \end{enumerate}

\section*{Acknowledgments}
We thank R. Skelton and I. Momcheva for very helpful discussions and support with the 3D-HST data.
Support for this work was provided by NASA through grant number HST-AR 13732 from the Space Telescope Science Institute, which is operated by AURA, Inc., under NASA contract NAS 5-26555.

The Bolshoi and MultiDark simulations have been performed within the Bolshoi project of the University of California High-Performance AstroComputing Center (UC-HiPACC) and were run at the NASA Ames Research Center. The MultiDark-Planck (MDPL) and the BigMD simulation suite have been performed in the Supermuc supercomputer at LRZ using time granted by PRACE.

PT acknowledges support from NASA ATP Grant NNX14AH35G. PT and MV acknowledge support through an MIT RSC award.  The Illustris simulation was run on the CURIE supercomputer at CEA/France as part of PRACE project RA0844, and the SuperMUC computer at the Leibniz Computing Centre, Germany, as part of project pr85je. The analysis presented in this paper was run on the Harvard Odyssey and CfA/ITC clusters and though allocation TG-AST150059 granted by the Extreme Science and Engineering Discovery Environment (XSEDE) supported by the NSF.

\bibliographystyle{apj_2}
\bibliography{references}

\label{lastpage}
\bsp
\end{document}